\documentclass[aps,pra,twocolumn,floats,nofootinbib]{revtex4-2} 
\usepackage{amsmath}
\usepackage{graphicx}
\usepackage{dcolumn}
\usepackage{bm}
\usepackage{amssymb}
\usepackage{latexsym}
\usepackage{color}
\usepackage{subfigure}
\usepackage{ulem}

\usepackage{cancel}

\begin{document}

\title{Hysteresis in the complex nonlinear refractive index of a homogeneous and isotropic medium}

\author{Igor Kuzmenko$^{1,2}$*, Y. B. Band$^{1,2,3}$*, Yshai Avishai$^{1,4}$, Marek Trippenbach$^{5}$}

\affiliation{
  $^1$Department of Physics,
  Ben-Gurion University of the Negev,
  Beer-Sheva 84105, Israel
  \\
  $^2$Department of Chemistry,
  Ben-Gurion University of the Negev,
  Beer-Sheva 84105, Israel
  \\
  $^3$The Ilse Katz Center for Nano-Science,
  Ben-Gurion University of the Negev,
  Beer-Sheva 84105, Israel
  \\
  $^4$Yukawa Institute for Theoretical Physics, Kyoto, Japan
  \\
  $^5$Faculty of Physics, University of Warsaw, ul. Pasteura 5, 02-093 Warszawa, Poland
  }

\begin{abstract}
We calculate the permittivity, $\epsilon(\omega)$, for a medium with a quadratic electro-optic effect, modeling it as a Duffing oscillator.  The nonlinear refractive index $n(\omega, E(\omega))$ and the nonlinear absorption coefficient $\alpha(\omega, E(\omega))$ exhibit hysteresis when the light intensity is varied [here $E(\omega)$ is the electric field strength at angular frequency $\omega$], and when the light frequency is varied.  $n(\omega, E(\omega))$ can be negative when the resonances in the permittivity and permeability are close to one another.
\end{abstract}

\maketitle

\def\thefootnote{*}\footnotetext{These authors contributed equally to this work.}

{\it Introduction}:  Negative-index metamaterials (NIMs) were first studied by Victor Veselago in 1967 \cite{Veselago_68}.  However, the first practical metamaterial with negative refractive index was not developed until 33 years later \cite{Shelby_01}.  NIMs can generate significant third harmonic generation (THG) when driven by strong electromagnetic fields, due to two primary factors: strong resonances that enhance local fields and the inclusion of materials with an inherent nonlinear optical response \cite{Xie-24}.

Here we consider a nonlinear generalization of the Drude-Lorentz model \cite{Drude-Lorentz, BF} using the the Duffing oscillator model \cite{Duffing} to investigate light propagation in isotropic homogeneous media under intense light conditions. Our aim is to generalize a previous study \cite{Band_24} which focused on the negative refraction in isotropic homogeneous media, by incorporating strong-field effects.  We demonstrate that the nonlinear refractive index $n(\omega, E(\omega))$ and the absorption coefficient $\alpha(\omega, E(\omega))$ (i.e., the complex nonlinear refractive index) of a homogeneous isotropic medium can exhibit hysteresis when the frequency and/or the intensity of the light are varied.  The real part of the complex nonlinear refractive index can be negative, however, for this to occur,  the resonance frequency of the permeability must be near the resonance frequency of the permittivity.

The hysteresis of light reflection and refraction at the boundary of a nonlinear medium has been well-studied (see references~\cite{Kaplan_77, Band_84, Afanasev_98, Lousse_03, Cai_22}). Here, however, we demonstrate that the complex nonlinear refractive index itself can exhibit hysteresis.

{\it The Duffing Model}:  Several studies have considered nonlinear Lorentz models, such as the Duffing oscillator \cite{Duffing,Stoker_50, Kazantseva-05, Kovacic-11}, to describe nonlinear optical dispersion, refraction and absorption \cite{Scalora-15, Varin-18, Varin-19, Xia-25, Chen_23}.  We use the Duffing equation, a nonlinear second-order differential equation, to model a bound electron (e.g., in an atom)  strongly driven by an electromagnetic field.  The  equation is
\begin{equation}   \label{eq:Duffing}
  \ddot{a}(t) +
  \gamma \, \dot{a}(t) + \omega_0^2 \, a(t) +
  \beta \, a^3(t) = \Gamma \, e^{- i \omega t} + \Gamma^* \, e^{i \omega t} ,
\end{equation}
where the real function $a(t)$ is the oscillator displacement coordinate at time $t$, $\dot{a}(t)$ is the velocity, $\ddot{a}(t)$ is the acceleration, $\omega_0$ is the resonance frequency of the (linear) oscillator, $\gamma$ is the decay rate, and $\beta$ is a nonlinearity parameter with units of s$^{-2}$cm$^{-2}$ in Gaussian units. The driving parameter, which has units of cm s$^{-2}$, is  given by $\Gamma = \frac{-e}{m} E$, where $e$ is the elementary charge, $m$ is the electron mass, and $E$ is the {\it complex} electric field strength of the electromagnetic radiation at frequency $\omega$.

We use an iterative method to treat the cubic term of the Duffing equation in order to solve Eq.~(\ref{eq:Duffing}). The first iteration includes only the fundamental frequency, the second iteration includes the fundamental frequency and the third harmonic frequency, etc.  Our method is somewhat similar to  the one described in Ref.~\cite{Stoker_50} which uses the iteration method originally employed by Duffing \cite{Duffing}.  Our iteration method, as opposed to other iteration methods \cite{Tisdell_19}, captures the multiple (i.e., three) solutions arising from the nonlinear term in the Duffing equation.

{\it Iteration method}:  
We are interested in cases where the nonlinear term is too large for perturbation theory to converge.  Therefore we have developed an iteration method which allows treating cases where the cubic term is not small.  In this section, we solve Eq.~(\ref{eq:Duffing}) using an iteration method.  The solution for $t \gg 2/\gamma$ is a periodic function with period $2 \pi / \omega$, and can be expanded in a Fourier series:
\begin{equation}   \label{eq:x-Fourier-expansion}
  a(t) \underset{t \gg 2/\gamma}{=}
  \sum_{n = 0}^{\infty}
  \big\{
    \beta^n \Gamma^{2 n + 1} \,
    {\mathcal A}_{2 n + 1}  \, e^{- i (2 n + 1) \omega t} +
    {\rm c.c.}
  \big\} .
\end{equation}
Here, $\beta^n \Gamma^{2 n + 1} {\mathcal A}_{2 n + 1}$ is the amplitude of the $(2 n +1)$th harmonic of the oscillator displacement coordinate.  In general, the coefficient ${\mathcal A}_{2 n + 1}$, has units of ${\text s}^{6 n + 2}$, and is proportional to the oscillator polarizability at frequency $(2 n + 1) \omega$.  In what follows, we describe the first iteration for $a(t)$.

The lowest order iteration in powers of $\beta \, \Gamma^2$, $a_1(t)$, can be obtained from Eq.~(\ref{eq:x-Fourier-expansion}) after neglecting all the harmonics and keeping only the fundamental frequency,
\begin{equation}   \label{eq:a_1-def}
  a_1(t) = \Gamma \, {\mathcal A}_{1} \, e^{- i \omega t} + {\rm c.c.}
\end{equation}
Substituting this solution into Eq.~(\ref{eq:Duffing}), and neglecting the third harmonic, yields
\begin{equation}   \label{eq:-for-A_1^1}
  \big\{ {\mathcal L}_{0}^{-1} (\omega) +3 \, \beta \, \big| \Gamma {\mathcal A}_1 \big|^{2} \big\} {\mathcal A}_1 = 1 ,
\end{equation}
where the linear Lorentzian operator is
\begin{equation}  \label{eq:L_0}
  {\mathcal L}_{0}(\omega) = \frac{- 1}{\omega^2 - \omega_0^2 + i \, \gamma \, \omega} .
\end{equation}
The solution for the amplitude ${\mathcal A}_1$ in Eq.~(\ref{eq:-for-A_1^1}) can be substituted into (\ref{eq:a_1-def}) to yield the first iteration for $a(t)$, i.e., $a_1(t)$.

The complex operator ${\mathcal L}_{0}(\omega)$ can be written in amplitude-phase notation as ${\mathcal L}_{0}(\omega) = | {\mathcal L}_{0}(\omega) | \, e^{i \theta_L (w)}$, where
\begin{eqnarray}  \label{eq:Labs}
  | {\mathcal L}_{0} (\omega) | &=&
  \frac{1}{\sqrt{(\omega^2 - \omega_0^2)^2 + \gamma^2 \omega^2}} ,
  \\
  \label{eq:Lphase}
  \theta_L(\omega) &=&
  \arctan \Big( \frac{\gamma \, \omega}{\omega_0^2 - \omega^2} \Big) .
\end{eqnarray}
${\mathcal A}_1$ can also be written in amplitude-phase notation as  ${\mathcal A}_1 = | {\mathcal A}_1 | \, e^{i \theta_A}$, where $| {\mathcal A}_1 |$ is found from the equation,
\begin{equation}   \label{eq:for-abs-A1}
  \big\{
    \big(
      \omega_0^2 - \omega^2 + 3 \, \beta \, \big| \Gamma \, {\mathcal A}_1 \big|^{2}
    \big)^{2} +
    \gamma^2 \omega^2
  \big\}
  \big| {\mathcal A}_1 \big|^{2} = 1 ,
\end{equation}
and the phase of ${\mathcal A}_1$ is $\theta_A = i \, \ln \big( | {\mathcal A}_1 | \,  \{ {\mathcal L}_{0}^{-1} (\omega) +3 \, \beta \, | \Gamma \, {\mathcal A}_1 |^{2} \} \big)$.  The complex coefficient ${\mathcal A}_1$ in Eq.~(\ref{eq:a_1-def}) can be found by solving cubic equation (\ref{eq:for-abs-A1}) for $| {\mathcal A}_1 |^{2}$ and substituting the solution into the expression for the phase.  The complex solution of the linear oscillator with $\beta = 0$, ${\mathcal A}_{1,0}$, is
\begin{equation}   \label{eqLA_1-0-res}
  {\mathcal A}_{1,0} = {\mathcal L}_{0}(\omega) .
\end{equation}

%
%
%
We denote the solutions of the cubic equation (\ref{eq:for-abs-A1}) by $|{\mathcal A}_{1, a}|^2$, $|{\mathcal A}_{1, b}|^2$ and $|{\mathcal A}_{1, c}|^2$.  Linear stability analysis shows that the solutions of Eq.~(\ref{eq:Duffing}), ${\mathcal A}_{1, a}$ and ${\mathcal A}_{1, b}$ are stable, and ${\mathcal A}_{1, c}$ is unstable.

The induced electric transition dipole moment of the oscillator, $d(t) \equiv - e a_1(t)$, can be written using Eq.~(\ref{eq:a_1-def}): $d(t) = \alpha_1(\omega) \, E \, e^{-i \omega t} + {\rm c.c.}$.  Here the oscillator polarizability at angular frequency $\omega$, $\alpha_1(\omega)$, is
\begin{equation}   \label{eq:alpha_1-vs-A_1}
  \alpha_1(\omega) = \frac{e^2}{m} \, {\mathcal A}_1 .
\end{equation}
Note that ${\mathcal A}_1$, hence also $\alpha_1$, is a nonlinear function of the amplitude of the electric field, $|E|$.

It is well known that the three solutions to the Duffing equation give rise to hysteresis in the nonlinear response for certain parameter values of the Duffing oscillator \cite{Duffing, Stoker_50, Kazantseva-05, Kovacic-11}.

\begin{figure}[htb]
\centering
  \includegraphics[width=0.9 \linewidth,angle=0] {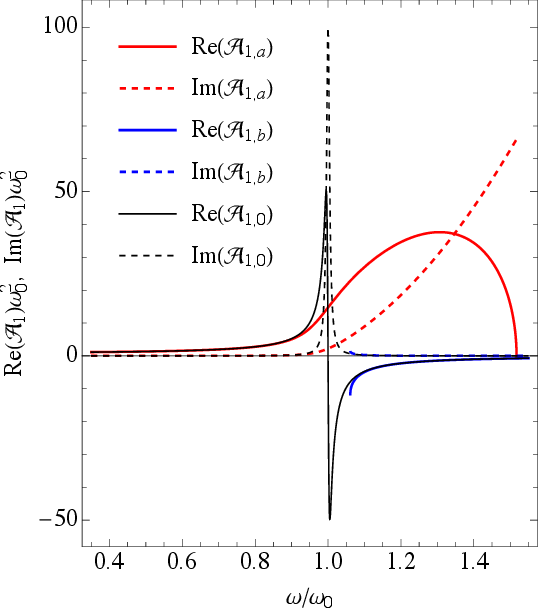}
\caption{\footnotesize The real and imaginary parts of the amplitudes ${\mathcal A}_{1, a}$,  red-solid and red-dashed curves respectively, and ${\mathcal A}_{1, b}$,  blue-solid and blue-dashed curves respectively, for $\gamma = 0.01 \, \omega_0$ and $\beta = 0.0001 \omega_0^6 / \Gamma^{2}$.  The black solid and dashed curves show the real and imaginary parts of ${\mathcal A}_{1,0}$ for the linear model (i.e., $\beta = 0$).}
\label{Fig:A1_Re-vs-w}
\end{figure}

Figure~\ref{Fig:A1_Re-vs-w} plots the real and imaginary parts of ${\mathcal A}_{1, a}$ and ${\mathcal A}_{1, b}$ versus $\omega$, and for comparison, the solution to the linear equation, ${\mathcal A}_{1,0}$.  The amplitude ${\mathcal A}_1$ as a function of the frequency $\omega$ manifests hysteresis behavior. If $\omega < \omega_0$ and $\omega_0 - \omega \gg \gamma$, then ${\mathcal A}_1 = {\mathcal A}_{1,a} \approx {\mathcal A}_{1,0}$, see the red and black curves in Fig.~\ref{Fig:A1_Re-vs-w}.  If the frequency undergoes a gradual increase, commencing at a point below a threshold of $\omega_1 = 1.0611 \, \omega_0$ yet remains below a threshold of $\omega_2 = 1.5175 \, \omega_0$, the amplitude is given by ${\mathcal A}_{1,a}$.  At $\omega = \omega_2$, ${\mathcal A}_1$ jumps from ${\mathcal A}_{1,a}$ to ${\mathcal A}_{1,b}$, and for $\omega - \omega_0 \gg \gamma$, ${\mathcal A}_{1,b} \approx {\mathcal A}_{1,0}$, see the blue and black curves in Fig.~\ref{Fig:A1_Re-vs-w}.  If the frequency undergoes a gradual decrease, commencing at a point above $\omega_2$ yet remains above $\omega_1$, the amplitude is given by ${\mathcal A}_{1,b}$.  At $\omega = \omega_1$, ${\mathcal A}_1$ jumps from ${\mathcal A}_{1,b}$ to ${\mathcal A}_{1,a}$.  Therefore, within the frequency interval $\omega_1 < \omega < \omega_2$ the amplitude ${\mathcal A}_1$ exhibits hysteresis behavior, meaning that it is influenced not only by the frequency $\omega$, but also by its history. This phenomenon is known as bistability, which occurs when two distinct steady states can coexist for a given frequency.

\begin{figure}[htb]
\centering
  \includegraphics[width=0.90 \linewidth,angle=0] {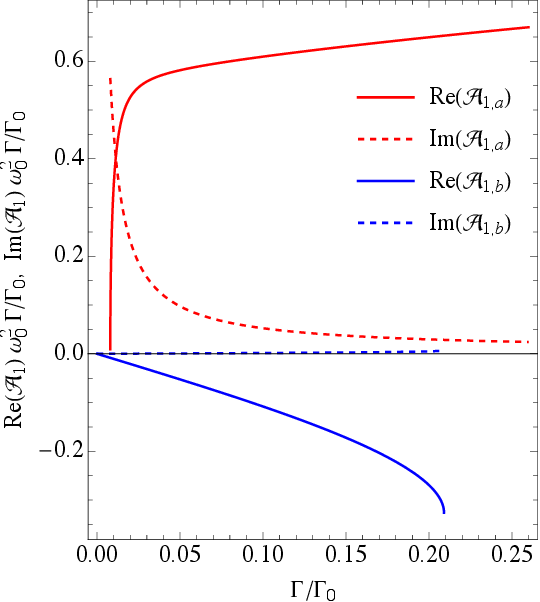}
\caption{\footnotesize The real and imaginary parts of the amplitudes ${\mathcal A}_{1, a}$ and ${\mathcal A}_{1, b}$, for $\gamma = 0.05 \, \omega_0$, 
and $\omega = 1.4 \, \omega_0$.}
\label{Fig:A1_Re-vs-G}
\end{figure}

Figure~\ref{Fig:A1_Re-vs-G} plots the real and imaginary parts of ${\mathcal A}_{1, a}$ and ${\mathcal A}_{1, b}$ versus $\Gamma$, where 
$\Gamma_0 = \tfrac{\omega_0^3}{\sqrt{\beta}}$, is the scale of $\Gamma$ used in the figure.  For $\Gamma < 0.0079 \, \Gamma_0$, Eq.~(\ref{eq:-for-A_1^1}) has a single solution, ${\mathcal A}_{1, b}$.  For $0.0079 \, \Gamma_0 < \Gamma < 0.2091 \, \Gamma_0$, there are two solutions, ${\mathcal A}_{1, a}$ and ${\mathcal A}_{1, b}$.  For $\Gamma > 0.2091 \, \Gamma_0$, Eq.~(\ref{eq:-for-A_1^1}) has a single solution, ${\mathcal A}_{1, a}$.  Note that varying $\beta$ results in varying of $\Gamma_0$, and the figure is invariant upon such a variation.

{\it Permittivity, Permeability and Refractive Index}:  
Let us consider propagation of a high intensity electromagnetic wave in an isotropic nonlinear medium.   
We model the medium as a set of 3D isotropic Duffing oscillators with number density $N$ distributed homogeneously in space.  The electric field of the electromagnetic wave can be written as
\begin{equation}
  {\bf E}({\bf r}, t) = \frac{1}{2} \, E_{\rm loc}({\bf r}) \, {\hat {\bf e}} \, e^{- i \omega t + i \Phi({\bf r})} + {\rm c.c.} ,
\end{equation}
where $E_{\rm loc}({\bf r})$ is the real amplitude of the local electric field with frequency $\omega$ (i.e., the external electric field plus the electric field created by the transition dipole moments), $\Phi({\bf r})$ is the spatially dependent phase, and ${\hat {\bf e}}$ is the unit vector in the direction of the polarization of the light.  The function $e^{i \Phi({\bf r})}$ varies rapidly with ${\bf r}$ and has a period of the light wavelength, and $E_{\rm loc}({\bf r})$ is a slowly-varying amplitude.

The induced electric transition dipole moment of a Duffing oscillator is
\begin{equation}
  {\bf d}({\bf r}, t) = d ({\bf r}) \,  e^{- i \omega t + i \Phi({\bf r})} \,  {\hat {\bf e}} ,
\end{equation}
where
\begin{equation}
  d ({\bf r}) = - e \, \Gamma \, {\mathcal A}_1 = \alpha_1(\omega) \, E_{\rm loc}({\bf r}) .
\end{equation}
${\mathcal A}_1$ is a solution of Eq.~(\ref{eq:-for-A_1^1}), and the complex polarizability of the Duffing oscillator $\alpha_1(\omega)$ is given in Eq.~(\ref{eq:alpha_1-vs-A_1}).  Multiplying the left- and right-hand sides of Eq.~(\ref{eq:-for-A_1^1}) by $- e \, \Gamma \equiv \frac{e^2}{m} \, E_{\rm loc}$, one obtains:
\begin{equation}   \label{eq:-for-d-vs-Eloc}
  \Big\{ \omega_0^2 - \omega^2 - i \, \gamma \, \omega +\frac{3 \, \beta}{e^2} \, \big| d_1 \big|^{2} \Big\} \, d_1 =
  \frac{e^2}{m} \, E_{\rm loc} .
\end{equation}
Solving this equation for the transition dipole moment $d_1$, one sees that it is dependent on the following variables: $\omega$, $\omega_0$, and $E_{\rm loc}$, i.e., $d_1(\omega, \omega_0, E_{\rm loc}) = \alpha_1(\omega, \omega_0, |E_{\rm loc}|) \, E_{\rm loc}({\bf r})$.  Note that the polarizability in Eq.~(\ref{eq:alpha_1-vs-A_1}) depends on the resonance frequency $\omega_0$ and is a nonlinear function of $|E_{\rm loc}|$.

In Gaussian units, the local electric field $E_{\rm loc}({\bf r})$ is expressed in terms of the external electric field $E_{\rm ext}({\bf r})$ and the polarization of the medium $P({\bf r}) = N d_1({\bf r})$ as \cite{Jackson_99}
\begin{equation}
  E_{\rm loc}({\bf r}) = E_{\rm ext}({\bf r}) + \frac{4 \pi}{3} \, N \, d({\bf r}) .
\end{equation}
Substituting this equation into Eq.~(\ref{eq:-for-d-vs-Eloc}) yields an equation of the same form, except that $E_{\rm loc}$ is replaced by $E_{\rm ext}$, and $\omega_0$ is replaced by the renormalized harmonic resonance frequency, $\tilde\omega_0$, where
\begin{equation}   \label{eq:omega_0-tilde}
  \tilde\omega_0 = \sqrt{\omega_0^2 - \frac{1}{3} \omega_p^2} ,
\end{equation}
and the plasma frequency squared, $\omega_p^2$, is
\begin{equation}   \label{eq:omega_p^2}
  \omega_p^2 = \frac{4 \pi N e^2}{m} .
\end{equation}
Equation~(\ref{eq:-for-d-vs-Eloc}) becomes
\begin{equation}   \label{eq:-for-d-vs-Eext}
  \Big\{ \tilde\omega_0^2 - \omega^2 - i \, \gamma \, \omega +\frac{3 \, \beta}{e^2} \, \big| d_1 \big|^{2} \Big\} \, d_1 =
  \frac{e^2}{m} \, E_{\rm ext} .
\end{equation}
The solution of Eq.~(\ref{eq:-for-d-vs-Eext}) yields,
\begin{equation}   \label{eq:d-vs-Eext}
  d_1(\omega, \omega_0, E_{\rm loc}) = \alpha_1(\omega, \tilde \omega_0, |E_{\rm ext}|) \, E_{\rm ext}({\bf r}) .
\end{equation}
Here $\alpha_1$ is still given by Eq.~(\ref{eq:alpha_1-vs-A_1}), except that $\omega_0$ is replaced by $\tilde \omega_0$ and $E_{\rm loc}$ is replaced by $E_{\rm ext}$.

The polarization of the medium, $P(\omega)$, can be written as
\begin{equation}   \label{eq:P-vs-Eext}
  P(\omega) = \chi(\omega) \, E_{\rm ext} ,
\end{equation}
where the electric susceptibility is given by
\begin{equation}   \label{eq:electric-susceptibility}
  \chi(\omega) = N \, \alpha_1(\omega, \tilde \omega_0, |E_{\rm ext}|) ,
\end{equation}
and the nonlinear permittivity $\epsilon(\omega)$ is \cite{Jackson_99}
\begin{eqnarray}   \label{eq:permittivity}
  \epsilon(\omega) &=&
  1 + 4 \pi \chi(\omega) =
  1 + 4 \pi N \, \alpha_1(\omega, \tilde \omega_0, |E_{\rm ext}|)
  \nonumber \\ &=&
  1 + \omega_p^2 \, {\mathcal A}_1(\omega, \tilde \omega_0, |E_{\rm ext}|) .
\end{eqnarray}
Here ${\mathcal A}_1(\omega, \tilde \omega_0, |E_{\rm ext}|)$ is the solution of Eq.~(\ref{eq:-for-A_1^1}) with $\omega_0$ replaced by $\tilde \omega_0$ and $E_{\rm loc}$ replaced by $E_{\rm ext}$.  This generalizes the Clausius-Mossotti relation for the linear permittivity to the nonlinear Duffing permittivity.  The permittivity for the linear Lorentz model, $\epsilon_L(\omega)$, is found from the Clausius-Mossotti relation \cite{Jackson_99},
\begin{equation}   \label{eq:Clausius-Mossotti}
  \frac{\epsilon_L(\omega) - 1}{\epsilon_L(\omega) + 2} = \frac{4 \pi N}{3} \, \alpha_L(\omega, \omega_0) ,
\end{equation}
where $\alpha_L(\omega, \omega_0)$ is
\begin{equation}   \label{eq:alpha_L}
  \alpha_L(\omega, \omega_0) = \frac{e^2}{m} \, {\mathcal L}_{0}(\omega, \omega_0) ,
\end{equation}
and the linear Lorentzian operator ${\mathcal L}_{0}(\omega, \omega_0)$ is given by Eq.~(\ref{eq:L_0}) (note that ${\mathcal L}_{0}$ depends on $\omega_0$).  The solution of Eq.~(\ref{eq:Clausius-Mossotti}) for $\epsilon_L(\omega)$ yields
\begin{equation}
  \epsilon_L(\omega) = 1 + \omega_p^2 \, {\mathcal L}_{0}(\omega, \tilde\omega_0) .
\end{equation}
The Clausius-Mossotti relation results in the replacement of $\omega_0$ by $\tilde \omega_0$.

\begin{figure}[htb]
\centering
  \includegraphics[width=0.90 \linewidth,angle=0] {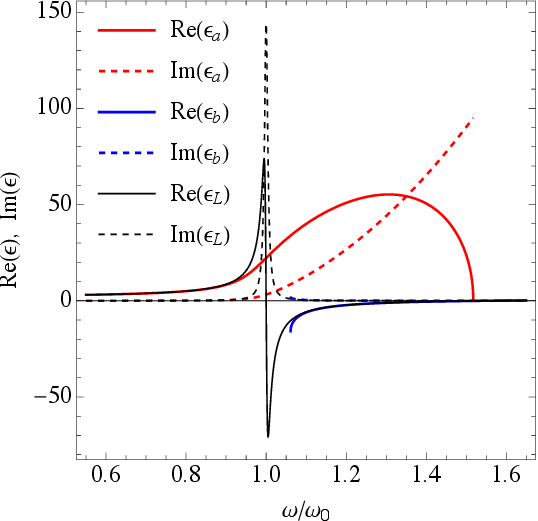}
\caption{\footnotesize The real and imaginary parts of the permittivities $\epsilon_a(\omega)$ and $\epsilon_b(\omega)$ for $\gamma = 0.01 \, \omega_0$, $\Gamma = 0.01 \, \Gamma_0$ and $\omega_p = 1.2 \, \omega_0$.}
\label{Fig:epsilon-Re-Im}
\end{figure}

Figure~\ref{Fig:epsilon-Re-Im} plots the real an imaginary parts of the permittivities $\epsilon_a(\omega)$ and $\epsilon_b(\omega)$.  For comparison, the figure also plots the real and imaginary parts of the Lorentzian permittivity $\epsilon_L(\omega)$.  The solution $\epsilon_a(\omega)$ ceases to exist for $\omega < \omega_2$, see Fig.~\ref{Fig:A1_Re-vs-w}. The real part of $\epsilon_a(\omega)$ is positive; it increases with increasing $\omega$ for $\omega < 1.3065 \, \omega_0$, and decreases in the region $1.3065 \, \omega_0 < \omega < \omega_2$.  The imaginary part of $\epsilon_a(\omega)$ increases with $\omega$.  Increasing the angular frequency above $\omega_2$, forces a transition from $\epsilon_a(\omega)$ to $\epsilon_b(\omega)$.  The solution $\epsilon_b(\omega)$ ceases to exist for $\omega > \omega_1$, see Fig.~\ref{Fig:A1_Re-vs-w}. The real part of $\epsilon_b(\omega)$ decreases with decreasing frequency. For $\omega > 1.56204 \, \omega_0$, $\epsilon_b(\omega)$ is positive, for $\omega_1 < \omega < 1.56204 \, \omega_0$, ${\rm Re}[\epsilon_b(\omega)]$ is negative; moreover, ${\rm Im}[\epsilon_b(\omega)]$ is small for this interval and for higher frequencies.  Decreasing the angular frequency below $\omega_1$, forces a transition from $\epsilon_b(\omega)$ to $\epsilon_a(\omega)$.  For $\omega_1 < \omega < \omega_2$, the permittivity of the medium depends on $\omega$, {\it and} on the history of the system. Consequently, the permittivity exhibits hysteresis.

A magnetic dipole transition with resonance frequency $\omega_{0,m}$, decay rate $\gamma_m$ and magnetic plasma frequency $\omega_{p,m}$ yields the magnetic permeability
\begin{equation}   \label{eq:mu}
  \mu(\omega) = 1 - \frac{\omega_{p,m}^{2}}{\omega^2 - \omega_{0, m}^{2} + i \gamma_m \omega} .
\end{equation}
Since magnetic transitions are usually smaller than electric transitions, we only consider linear magnetic permeability.

The complex refractive index can be written as
\begin{equation}   \label{eq:n_def}
  n(\omega) = \sqrt{|\epsilon(\omega)| \, |\mu(\omega)|} \, e^{i [\theta_{\epsilon}(\omega) + \theta_{\mu}(\omega)]/2} ,
\end{equation}
where $|\epsilon(\omega)|$ and $\theta_{\epsilon}(\omega)$ are the absolute value and the complex phase of the complex permittivity $\epsilon(\omega)$, and $|\mu(\omega)|$ and $\theta_{\mu}(\omega)$ are the absolute value and the complex phase of $\mu(\omega)$.  Substituting either $\epsilon_a(\omega)$ or $\epsilon_b(\omega)$ for $\epsilon(\omega)$ in Eq.~(\ref{eq:n_def}) results in the refractive indices $n_a(\omega)$ or $n_b(\omega)$ for the Duffing model.  Substituting $\epsilon_L(\omega)$  for $\epsilon(\omega)$ in Eq.~(\ref{eq:n_def})  results in the Drude-Lorentz linear refractive index, $n_L(\omega)$.  Note that the absorption coefficient is given by $\alpha(\omega, E(\omega)) = \omega \, {\mathrm{Im}}[n(\omega)]/c$.

\begin{figure}[htb]
\centering
  \includegraphics[width=0.90 \linewidth,angle=0] {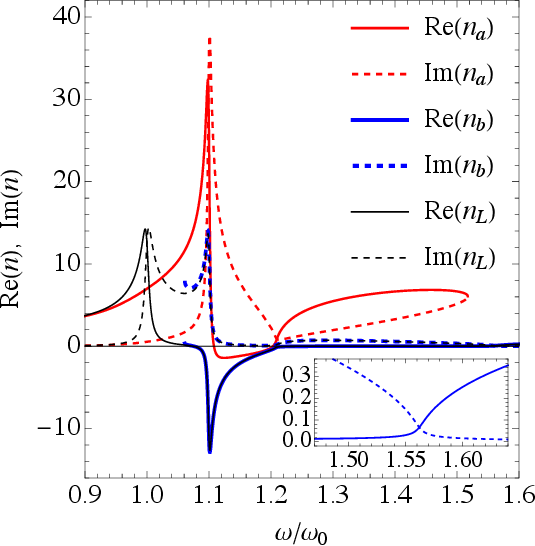}
\caption{\footnotesize The real (solid) and imaginary (dashed) parts of the refractive indices $n_a(\omega)$ (red) and $n_b(\omega)$ (blue) are plotted versus frequency using the parameters $\gamma = 0.01 \, \omega_0$, $\Gamma = 0.01 \, \Gamma_0$, $\omega_p = 1.2 \, \omega_0$, $\gamma_m = 0.005 \, \omega_0$, $\omega_{p,m} = 0.5 \, \omega_0$ and $\omega_{0,m} = 1.1 \, \omega_0$.  The solid and dashed black curves show the real and imaginary parts of the refractive index $n_L(\omega)$ for the Lorentz model. The inset is a zoom of the real and imaginary parts of $n_b(\omega)$ for large $\omega$. }
\label{Fig.na_nb_vs_w}
\end{figure}

Figure~\ref{Fig.na_nb_vs_w} shows the real and imaginary parts of the refractive indices $n_a(\omega)$ and $n_b(\omega)$ versus $\omega$.  For comparison, the real and imaginary parts of the refractive index $n_L(\omega)$ versus $\omega$ are also plotted in Fig.~\ref{Fig.na_nb_vs_w}.  For $\omega < \omega_1 = 1.0611 \, \omega_0$, only state $a$ exists, and for $\omega > \omega_2 = 1.5175 \, \omega_0$, only state $b$ exists.  For $\omega < \omega_0$ and $\omega_0 - \omega \gg \gamma$, $n_a(\omega) \approx n_L(\omega)$, and the nonlinear effects are small.    For $\omega > \omega_0$ and $\omega - \omega_0 \gg \gamma$, $n_b(\omega) \approx n_L(\omega)$.  The real part of $n_a(\omega)$ is negative in the interval $1.1 \, \omega_{0} < \omega < 1.24 \, \omega_{0}$, but the imaginary part is large.  This is contrary to the double negative concept introduced in \cite{Veselago_68} since Re[$\epsilon_a(\omega)$] is positive within this interval, see Fig.~\ref{Fig:epsilon-Re-Im}.  Increasing the angular frequency above $\omega_2$ forces a transition from $n_a(\omega)$ to $n_b(\omega)$.  For $1.076 \, \omega_0 < \omega < 1.246 \, \omega_0$, the real part of $n_b(\omega)$ is negative, and the imaginary part is relatively small.  Decreasing the angular frequency below $\omega_1$, forces a transition from $n_b(\omega)$ to $n_a(\omega)$.  For $\omega_1 < \omega < \omega_2$, the refractive index of the medium depends on $\omega$ {\it and on the history of the system}, i.e., the refractive index exhibits hysteresis.  In this frequency interval, $n_b(\omega) \approx n_L(\omega)$, however, $n_a(\omega)$ is far from $n_L(\omega)$, therefore the nonlinearity is a salient factor when the medium is in state $a$.

\begin{figure}[htb]
\centering
  \includegraphics[width=0.90 \linewidth,angle=0] {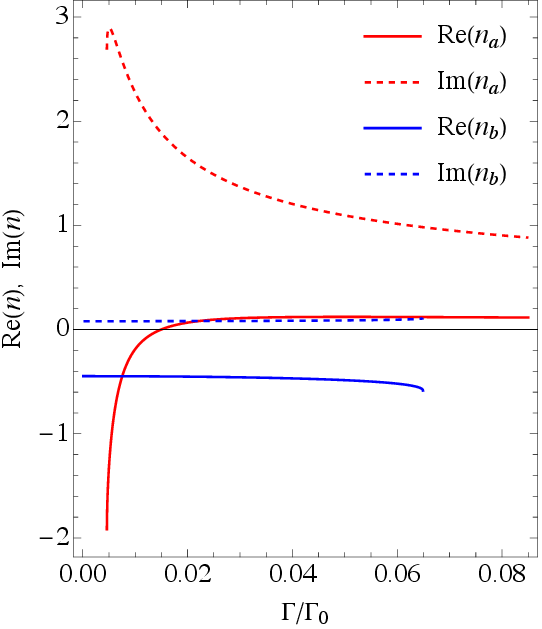}
\caption{\footnotesize The real and imaginary parts of the refractive indices $n_a(\omega)$ and $n_b(\omega)$ versus field strength for $\omega = 1.2 \, \omega_0$, $\gamma = 0.01 \, \omega_0$, $\omega_p = 1.2 \, \omega_0$, $\gamma_m = 0.005 \, \omega_0$, $\omega_{p,m} = 0.5 \, \omega_0$ and $\omega_{0,m} = 1.1 \, \omega_0$. }
\label{Fig.na_nb_vs_G}
\end{figure}

Figure \ref{Fig.na_nb_vs_G} shows real and imaginary parts of the refractive indices $n_a(\omega)$ and $n_b(\omega)$ versus $\Gamma$ for $\omega = 1.2 \, \omega_0$.  For $\Gamma < 0.005 \Gamma_0$ only state $b$ exists, for $\Gamma > 0.065 \Gamma_0$ only state $a$ exists, and in between both states exist.  The refractive index of the system depends on the intensity of the field as well as on the history of the intensity, i.e., the refractive index exhibits hysteresis.  Starting at low intensity, the real part of the refractive index $n_b(\omega)$ is negative, and the imaginary part is small.  Increasing the intensity so that $\Gamma > 0.065 \, \Gamma_0$, the system jumps to $n_a(\omega)$, whose real part is positive for $\Gamma > 0.065 \, \Gamma_0$.  Lowering the intensity so $\Gamma < 0.065 \, \Gamma_0$, the system remains in state $a$, and the refractive index becomes negative below $\Gamma = 0.015 \, \Gamma_0$.  Upon continuing to lower the intensity, the system jumps to state $b$ at $\Gamma = 0.005 \, \Gamma_0$ (where the red curves cease to exist), and the absorption falls dramatically to a small value.

{\it Summary and Conclusion}: Using the Duffing oscillator model for the electric transition dipole moment and the Drude-Lorentz model for the magnetic transition moment, we have shown that for light of sufficient intensity that propagates in an optically homogeneous and isotropic medium, the nonlinear refractive index $n(\omega, E(\omega))$ and nonlinear absorption coefficient $\alpha(\omega, E(\omega))$ exhibit hysteresis when the intensity and frequency of the light are varied.  The nonlinear refractive index can be negative in certain frequency regions and certain intensity regions (this is a generalization of the results in Ref.~\cite{Band_24} to the nonlinear regime).  Future work will discuss the properties of the time-averaged Poynting vector (i.e., the time-averaged energy flux density), energy density, power dissipation density, phase velocity, and group velocity in the nonlinear regime. We will also generalize to the chiral case.


\end{document}